\documentclass{elsart3-1}

\usepackage{amssymb}
\usepackage{epsfig}

\usepackage[english,francais]{babel}

\newtheorem{e-proposition}[theorem]{Proposition}

\newtheorem{e-definition}[theorem]{Definition\rm}

\renewcommand{\theequation}{\arabic{equation}}
\def\og{\leavevmode\raise.3ex\hbox{$\scriptscriptstyle\langle\!\langle$~}}
\def\fg{\leavevmode\raise.3ex\hbox{~$\!\scriptscriptstyle\,\rangle\!\rangle$}}

\newcommand{\be}{\begin{equation}}
\newcommand{\ee}{\end{equation}}
\newcommand{\bea}{\begin{eqnarray}}
\newcommand{\eea}{\end{eqnarray}}

\begin{document}
\centerline{{Future flavour physics experiments. 
Exp\'eriences futures de physique de la saveur}}
\begin{frontmatter}


\selectlanguage{english}
\title{Flavour Violation in charged leptons: Present and Future}


\selectlanguage{english}
\author[authorlabel1]{A. Abada} \ead{asmaa.abada@th.u-psud.fr}\\

\address[authorlabel1]{Laboratoire de Physique Th\'eorique, CNRS -- UMR 8627, 
Universit\'e de Paris-Sud 11, F-91405 Orsay Cedex, France}
\begin{abstract}
In the absence of a fundamental principle preventing charged 
lepton flavour violation, one expects that extensions of the 
Standard Model  accommodating
neutrino masses and mixings should also allow for charged lepton flavour
violating processes such as $\ell_i \to \ell_j\gamma$,  $\ell_i \to \ell_j \ell_k \ell_m $ and $\mu - e$ conversion in nuclei, for which 
 the rates  depend in general on the mechanism of neutrino mass generation. In addition to low-energy experiments, there are also searches for lepton flavour violation at colliders, where new physics can be directly  probed  through flavour violating production and/or decays of heavy states.
In a model independent way, we briefly use effective operators responsible for these processes to derive information about the underlying framework of new physics.  We then consider some specific classes of models (supersymmetry, extra dimensions, grand unified theories) that account for rich scenarios of charged lepton flavour violation. We also comment on the r\^ole of 
charged lepton flavour violation in disentangling models of new physics. \\
{\it To cite this article: A. Abada, C.R. Physique XX (2010).}

\vskip 0.5\baselineskip

\selectlanguage{francais}
\noindent{\bf R\'esum\'e}
\vskip 0.5\baselineskip
\noindent
{\bf Violation de la saveur dans le secteur des leptons
  charg\'es : Pr\'esent et Futur.}\\  
\vskip 0.5\baselineskip
\noindent  En l'absence de tout principe fondamental qui interdirait la non-conservation de la saveur dans le secteur des leptons charg\'es, on s'attend que les extensions du Mod\`ele Standard  qui expliquent les masses et m\'elanges des neutrinos autorisent \'egalement la violation de la saveur \`a travers des processus tels que $\ell_i \to \ell_j\gamma$, $\ell_i \to \ell_j \ell_k \ell_m $   et  la conversion  $\mu - e$ dans les noyaux, 
dont les taux d\'ependent  en g\'en\'eral  
du m\'ecanisme qui engendre des masses de
 neutrinos. 
 On peut aussi rechercher des signes de cette violation 
directement aux collisionneurs, \`a travers des processus de production/d\'esintegration de particules lourdes. D'une fa\c con ind\'ependante de tout mod\`ele, nous montrons 
comment obtenir des informations sur la physique sous-jacente en utilisant des op\'erateurs effectifs. 
Enfin, nous commenterons sur le r\^ole de la violation de la saveur leptonique dans la tri entre des classes de mod\`eles (supersym\'etrie, dimensions suppl\'ementaires, th\'eories grand-unifi\'ees) en ce qui concerne  la saveur leptonique.
\\
{\it Pour citer cet article~: A. Abada, C. R. Physique XX (2010).}
 \vskip 0.5\baselineskip 
\noindent{\small{\it Mots-cl\'es~:}  Violation de la Saveur
  Leptonique~;  Physique des Neutrinos~; Th\'eories Effectives
  ~; Nouvelle Physique}} 
\end{abstract}
\end{frontmatter}

\selectlanguage{english}

\section{Introduction}
\label{intro}
\begin{figure}[h!]
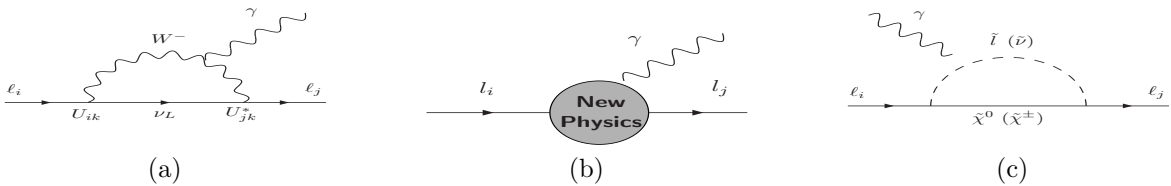

\begin{center}
\begin{tabular}{ccc}
\raisebox{3mm}{\epsfig{file=LFVSM.epsi, height=15mm, width=43mm}}
\hspace*{5mm} & \hspace*{5mm}
\raisebox{0mm}{\epsfig{file=lj.bbox.li.NP.epsi,  height=15mm,width=43mm}}
\hspace*{5mm} &\hspace*{5mm}
\raisebox{2.7mm}{\epsfig{file=LFV01.epsi,  height=15mm,width=43mm}} \vspace*{-2mm}\\
\vspace*{-2mm}(a) \hspace*{5mm}&\hspace*{5mm} (b) \hspace*{5mm}&\hspace*{5mm} (c)
\end{tabular}
\caption{
Radiative decays $\ell_i \to \ell_j \gamma$: (a) in the SM ($m_\nu\neq 0$), in new physics models (b) and in SUSY models (c). D\'esint\'egrations radiatives $\ell_i \to \ell_j \gamma$ dans les cadres : (a) du Mod\`ele Standard ($m_\nu\neq 0$), de nouvelle physique (b) et de la sypersym\'etrique (c).  }\label{LFVBSM}
\end{center}
\end{figure}

Neutrino ($\nu$) oscillation experiments provide indisputable evidence for flavour
violation in the neutral lepton sector.  
One expects that extensions of the 
Standard Model (SM) accommodating
neutrino masses ($m_\nu$) and mixings should also allow for charged 
lepton flavour violation (cLFV) (for a review, see 
Ref.~\cite{Raidal:2008jk} and references therein). 
In the SM, as it was originally formulated (massless $\nu$), 
cLFV processes are forbidden. 
With massive $\nu$s (no assumption being made on the 
mechanism of $\nu$ mass generation), 
cLFV processes are suppressed by 
the tiny $\nu$ masses (GIM mechanism). 
For example, the branching ratio (BR) of the $\mu \to e \gamma$ decay, Figure \ref{LFVBSM}(a), one has
\begin{equation}\label{eq:SM:BRmue}
\mathrm{BR}(\mu \to e \gamma) = \frac{3 \alpha}{32 \pi} \left|   
\sum_i U^*_{\mu i} U_{e i} \frac{m^2_{\nu_i}}{M_W^2}
\right|^2\,,
\end{equation}
and using known oscillation parameters 
($U=U_{\mathrm{MNS}}$ being the leptonic mixing matrix)~\cite{PDG}, one finds 
BR$(\mu \to e \gamma)\lesssim 10^{-54}$, 
thus clearly unaccessible to present and future experiments!
In contrast with FV in the hadronic sector, where
phenomena such as neutral meson mixings and rare decays can be
successfully explained by the SM, 
cLFV undisputably signals the presence of new physics (NP).  Indeed, 
the additional  particle content and new flavour dynamics present 
in many extensions of the SM may give contributions to 
cLFV processes such as radiative 
(e.g. $\mu \to e\gamma$) and three-body decays  
(e.g. $\mu \to eee$), so that 
the observation of such processes would provide an unambiguous 
signal of NP, see Figure \ref{LFVBSM}(b).

The quest for new physics  is currently being pursued along different
avenues:  
although high-energy colliders like the LHC are the ideal tools to directly  
search the particle content of the SM extension, low-energy
experiments indirectly probe the NP model through its
contributions to several observables (among which the muon anomalous magnetic
moment, electric dipole moments, cLFV observables, B-physics, etc.).
Thus, if observable, cLFV is expected to play a crucial r\^ole in  
successfully (or even partially) reconstructing the underlying
framework of new physics. 

In the SM, lepton number is an accidental symmetry due to the gauge
group and particle content. The generation of $\nu$ masses 
exclusively using the SM field content requires adding to the 
SM Lagrangian non-renormalisable operators of dimension 5 (or
higher), that break lepton number. The unique dimension 5 operator
that can be built with the SM fields is the Weinberg operator, which
is responsible for the Majorana mass of the neutrinos. 
In fact, any model of new physics that violates $B-L$ will give rise
to the Weinberg operator at low energies (when the heavy fields are 
integrated out). Among the dimension 6 operators (second order in
an $1/ M$ expansion, $M$ being the scale of new physics), one finds 
a four-fermion operator responsible for cLFV processes.   
The breaking of lepton number, as required by Majorana $\nu$ masses, then 
provides a natural link between neutrino mass generation and cLFV.  
\section{Current experimental searches and future prospects}
So far, no signal of cLFV has been observed. 
Although the searches for rare leptonic decays have been an
important part of the experimental program for several decades, 
these searches gained a renewed momentum with the confirmation of neutral 
LFV (neutrino oscillations). Currently, 
the search for manifestations of charged LFV constitutes the goal of 
several
experiments~\cite{Brooks:1999pu}-\cite{Hayasaka:2007vc}, 
exclusively dedicated to look for signals of  rare lepton decay 
processes. Equally interesting LFV observables are $\mu-e$ conversions in heavy
nuclei: although significant improvements are expected regarding the
experimental sensitivity to $\mu \to e \gamma$ ($ <
10^{-13}$~\cite{Kiselev:2009zz}), the most challenging experimental
prospects arise for the CR($\mu-e$) in heavy nuclei such as titanium
or gold~\cite{Glenzinski:2010zz,Cui:2009zz}. The possibility of improving the sensitivities to values as
low as $ \sim 10^{-18}$ renders this observable an extremely powerful
probe of LFV in the muon-electron sector. 
In the presence of CP violation, one can further have T- and P-odd 
asymmetries in cLFV decays and contributions to lepton electric dipole
moments. 
In Table~\ref{table:LFV:bounds}, 
we  briefly survey some of the current bounds for the different cLFV
processes, as well as the future experimental sensitivity. 
{\small
\begin{table}[h!]
\begin{center}
\begin{tabular}{cc}
\begin{tabular}{|l|c r|c r|}
\hline
LFV process & Present bound & & Future sensitivity & \\
\hline
BR($\mu \to e \gamma$) & $1.2 \times 10^{-11}$&
\cite{PDG}&$10^{-13} $& \cite{Kiselev:2009zz} \\ 
BR($\tau \to e \gamma$) &$1.1 \times 10^{-7}$ & \cite{Aubert:2005wa}&
$ 10^{-9}$& \cite{Bona:2007qt} \\ 
BR($\tau \to \mu \gamma$) & $4.5 \times 10^{-8}$&
\cite{Hayasaka:2007vc}&$ 10^{-9}$ & \cite{Bona:2007qt} \\ 
\hline
BR($\mu \to 3 e $) &$1.0 \times 10^{-12}$ &
\cite{PDG}&
&  \\ 
BR($\tau \to 3 e $) & $3.6 \times 10^{-8}$& \cite{PDG}&$2
\times 10^{-10}$ & \cite{Bona:2007qt} \\ 
BR($\tau \to 3 \mu$) & $3.2 \times 10^{-8}$& \cite{PDG}&$2
\times 10^{-10}$ & \cite{Bona:2007qt} \\ \hline
\end{tabular}
\hspace*{3mm}&\hspace*{3mm}
\begin{tabular}{|l|c r|c r|}
\hline
LFV process & Present bound & & Future sensitivity & \\
\hline
CR($\mu-e$, Ti) & $4.3 \times 10^{-12}$& \cite{PDG}&
${\mathcal{O}}(10^{-16 (-18)})$ &
\cite{Glenzinski:2010zz}~(\cite{Cui:2009zz}) \\ 
CR($\mu-e$, Au) & $7 \times 10^{-13}$& \cite{PDG}& 
& 
 \\
\hline 
CR($\mu-e$, Al) && & ${\mathcal{O}}(10^{-16})$&  \cite{Cui:2009zz}\\\hline
\end{tabular}
\end{tabular}
\end{center}
\caption{Present bounds and future sensitivities for several LFV
  observables. \newline
  Limites actuelles et sensibilit\'es futures pour diverses observables violant la saveur leptonique.}
\label{table:LFV:bounds}
\end{table}
}

In addition to low-energy experiments, there are also searches for cLFV
at high-energy: the presence of new flavour violating physics 
can be directly signaled via 
the LFV production and/or  decays of heavy states 
(which must be  nevertheless sufficiently light to be produced at present
colliders (Tevatron, LHC)).   
\section{LFV within effective theories}
In the absence of direct evidence for new physics, 
one can either consider well motivated models at high-energy scales
and study their impact on low-energy observables (the top-down
approach), or adopt an effective approach, which consists in
studying in a model-independent way the impact of the potentially
heavy new degrees of freedom. 
When integrated out, these heavy fields give rise to 
non-renormalisable effective operators that may
induce contributions to several observables (among which cLFV).
\subsection{Neutrino mass generation}
There are typically three possible ways to generate non-vanishing
$m_\nu$. Firstly, $\nu $ masses can be  
generated radiatively through higher order loop corrections 
(as in the case of R-parity violating supersymmetric (SUSY) models~\cite{Barbier:2004ez}, 
variations of Zee model, etc.). 
The second option consists in invoking a geometrical suppression mechanism: 
the SM fields can be localised on a four-dimensional brane, while the
extra heavy fields responsible for $m_\nu\ne 0$ live in
the bulk. Their couplings to the SM fields are small
due to geometrical suppression factors of large extraD, thus
inducing small $m_\nu$. Finally, the most elegant mechanism is  perhaps the so-called seesaw mechanism, 
which generates small $m_\nu$ at tree level via the
exchange of extra heavy fields. 
Here we will focus on the different realisations of this
mechanism~\cite{Abada:2007ux}.\\
It can be shown that one can only have  three
types of basic  seesaw mechanisms, depending on the nature of the new
heavy fields: right-handed neutrinos (type I)~\cite{seesaw:I}, 
heavy scalars (type II)~\cite{seesaw:II} or 
fermionic triplets (type III)~\cite{seesaw:III}, as depicted in
Figure~\ref{fig:seesawI-III}. It is important to notice that all 
these mechanisms can be embedded into larger frameworks such as grand unified theories (GUTs), SUSY and  extradimensions (extraD). Moreover,
one can have simultaneous combinations of different seesaw types
(e.g. I and III), or inverse seesaw.  
Hereafter, we will focus on the effective operators 
characteristic to each realisation.  
\begin{figure}[h!]
\begin{center}
\begin{tabular}{ccc}
\raisebox{3mm}{\epsfig{file=fig.type.I.epsi, height=17mm, width=38mm}}
\hspace*{5mm} & \hspace*{5mm}
\raisebox{3mm}{\epsfig{file=fig.type.II.epsi,  height=17mm, width=38mm}}
\hspace*{5mm} &\hspace*{5mm}
\raisebox{0mm}{\epsfig{file=fig.type.III.epsi,    height=19mm, width=25mm}} \vspace*{-3mm}\\
 \vspace*{-3mm}
(a) \hspace*{5mm}&\hspace*{5mm} (b) \hspace*{5mm}&\hspace*{5mm} (c)
\end{tabular}
\caption{
Seesaw mechanisms: (a) singlet fermion,  (b) triplet fermion
and (c) triplet scalar exchange. \newline
Les m\'ecanismes de la balan\c coire : \'echange de  (a) singlet fermionique,  (b) triplet fermionique
et (c) triplet scalar.} 
\label{fig:seesawI-III}
\end{center}
\end{figure}

\subsection{Effective approach}
In the effective approach, the higher dimensional operators are
obtained when expanding the heavy field propagators in $1/M$, $M$
being their mass, i.e. the scale of physics beyond the SM. In the case of a type I or III seesaw, 
the heavy fermionic propagator is expanded as 
$\frac{1}{ D\!\!\!\!/ \ -M}\,\sim\, -{\frac{1}{M}}\,-\, \frac{1}{M}\, D\!\!\!\!/
\frac{1}{M}+\cdots$
The first term scales as $1/M$, thus inducing a $d=5$ scalar
operator, which flips chirality and generates a $\nu$ mass term. 
The second term ($\sim 1/M^2$) preserves chirality and
induces a correction to the kinetic terms of the light fields, 
since it is proportional to the covariant derivative. 
The coefficients that weigh the $d=6$ operator,
$ c^{d=6}\propto   {1\over M^2}$, are
suppressed compared to those associated to the $d=5$ operator, 
$c^{d=5}\propto  {  {1\over M}}$.  
The situation is different in the case of heavy scalar triplets,
since the scalar propagator expands as 
$\frac{1}{D^2 -M^2}\,\sim\, -{\frac{1}{M^2}}\,-\, \frac{D^2}{M^4} +\ \cdots$, 
implying that the $d=5$ operator already scales as $1/M^2$. 

Independently of the model, the only possible $d=5$
operator is the Weinberg operator~\cite{Weinberg:1979sa}, 
$
\delta{\mathcal L}^{d=5} = \frac{1}{2}\, c_{\alpha \beta}^{d=5} \,
\left( \overline{\ell_L^c}_{\alpha} \tilde \phi^* \right) \left(
\tilde \phi^\dagger \, {\ell_L}_{ \beta} \right) + {\rm h.c}.\, ,
$
where $\ell_L$ stands for the lepton doublets and 
and $\tilde \phi$ is related to the SM Higgs doublet. 
The coefficient $c^{d=5}$ is a matrix of 
inverse mass dimension, which is not invariant under  $B-L$,
 and is thus a source of Majorana $\nu$ masses. 
In the case of the type I and III seesaws,
$c^{d=5}=Y_{N}^{T}\frac{1}{M_{N}}Y_{N}$, $Y_{N}$ being
the Yukawa couplings to the Higgs field 
and $M_N$ the heavy fermion masses. 
Accommodating $\nu$ data with natural coefficients ($c^{d=5} \sim {\mathcal{O}}(1)$), implies $ M_N\sim 10^{15}$ GeV,
intriguingly close to GUT scale. However,  
the scale of NP can be lowered to $\sim 1 $ TeV 
if one allows for couplings as small as the charged lepton ones.
In the case of a scalar triplet, $c^{d=5} \propto Y_\Delta
\mu_\Delta/M^2_\Delta$: the scale $\mu_\Delta$ can be directly related
to the smallness of $m_\nu$ thus allowing to have $M_\Delta \sim $ TeV
with natural Yukawa couplings.

More importantly, and since such a $d=5$ operator is
characteristic to all models with Majorana $\nu$, 
the coefficient $c^{d=5}$ does not allow to
discriminate among the different models. In order to do so, 
one must 
either produce the heavy mediators or call upon the low-energy effects
of the different  $d=6$ operators. There is a large number   
of such operators~\cite{Buchmuller:1985jz} but here we will only 
focus on those inducing cLFV processes~\footnote{We will not address 
here issues such as electroweak precision observables and 
unitarity violation.}. On Table~\ref{table:operators} we list the  $d=6$ 
cLFV operators as well as the corresponding coefficient 
for each type of seesaw and, for comparison, the $d=5$ coefficient.  
\begin{table}[!h]
\begin{center}
\begin{tabular}{|c||c|c|c|}
\hline
& \multicolumn{3}{c|}{Effective Lagrangian
  $\mathcal{L}_{eff}=c_{i}\mathcal{O}_{i} $} \rule[-5 pt]{0pt}{18
  pt}\\ 
\cline{2-4}
Model & $c^{d=5}$ & $c^{d=6}$ & $\mathcal{O}^{d=6}$  \rule[-5
pt]{0pt}{18 pt}\\ 
\hline 
\hline
$\begin{array}{c}
\mathrm{Fermionic \,Singlet} \vspace*{-1mm}\\
\mathrm{(type \,I)}
\end{array}$
& $Y_{N}^{T}\frac{1}{M_{N}}Y_{N}$ &
$\left(Y_{N}^{\dagger}\frac{1}{M_{N}^\dagger}\frac{1}{
    M_N}Y_{N}\right)_{\alpha\beta}$ &
$\left(\overline{\ell_{L\alpha}}\widetilde{\phi}\right)i\partial\!\!\!/
\left(\widetilde{\phi}^{\dagger}\ell_{L\beta}\right)$ 
\rule[-14 pt]{0pt}{34 pt}\\ 
\hline
$\begin{array}{c}
\mathrm{Scalar \,Triplet } \vspace*{-1mm}\\
\mathrm{(type \,II)}
\end{array}$
 & $4Y_{\Delta}\frac{\mu_{\Delta}}{M_{\Delta}^{2}}$ &
$\frac{1}{ M_\Delta^{2}}Y_{\Delta \alpha\beta}Y_{\Delta
  \gamma\delta}^{\dagger}$ &$
\left(\overline{\widetilde{\ell_{L\alpha}}}
\overrightarrow{\tau}\ell_{L\beta}\right)\left(\overline{\ell_{L\gamma}}
\overrightarrow{\tau}
  \widetilde{\ell_{L\delta}}\right)$
\rule[-14 pt]{0pt}{34 pt}\\ 
\hline
$\begin{array}{c}
\mathrm{Fermionic \,Triplet} \vspace*{-1mm}\\
\mathrm{(type \,III)}
\end{array}$
 & $Y_{\Sigma}^{T}\frac{1}{M_{\Sigma}}Y_{\Sigma}$ &
$\left(Y_{\Sigma}^{\dagger}\frac{1}{M_{\Sigma}^\dagger}\frac{1}{
    M_\Sigma}Y_{\Sigma}\right)_{\alpha\beta}$ &
$\left(\overline{\ell_{L\alpha}}\overrightarrow{\tau}
\widetilde{\phi}\right)iD\!\!\!\!/\left(
\widetilde{\phi}^{\dagger}\overrightarrow{\tau}\ell_{L\beta}\right)$ 
\rule[-14 pt]{0pt}{34 pt}\\ 
\hline
\end{tabular}
\end{center}
\caption{Dimension 6 operators (and their coefficients) responsible for cLFV, also 
displaying the corresponding $d=5$ coefficient. \newline
Op\'erateurs de dimension 6 (et leurs coefficients) des processus cLFV ainsi que les coefficients $d=5$ correspondants. }
\label{table:operators}
\end{table}\\
From a symmetry point of view, it is natural to have large $c^{d=6}$ 
coefficients, since the $d=6$ operators preserve $B-L$, 
in contrast with the $d=5$ operator. For example, 
in the type II seesaw, the dimensionful $\mu_\Delta$ coefficient, 
which is directly related to the smallness of $m_\nu$,
does not affect the dimension 6 operator. 
However, decoupling the $d=5$ and $d=6$ coefficients is not
possible in the fermionic seesaw, see   
Table~\ref{table:operators}. 
\subsection{Charged lepton flavour violation}
In the effective approach, the observables can be written in
terms of effective parameters, encoding the flavour mixing
generated by the model, which remains valid  up to a scale $\Lambda$. For example, in models onto which a type II seesaw is embedded ($\Lambda=M_\Delta$),
the BRs for radiative and three body decays read 
\begin{equation}\label{eq:BR:eff}
\frac{\textrm{BR}(l_{i}\rightarrow l_{j}\gamma)}
{\textrm{BR}(l_i\rightarrow l_{j} \nu_i \bar{\nu}_j)}
={\alpha \over 48 \pi} 
{25\over 16} \frac{\left|{\Sigma_k}{Y_\Delta}_{l_kl_{i}}^{\dagger}
{Y_\Delta}_{l_{j}l_k}\right|^{2}}{G_{F}^{2}\Lambda^4}\, 
\,;\quad
\textrm{BR}(\ell_l^-\rightarrow l_i^+ l_j^- l_j^-)
= {1\over  { G_F^2\Lambda}^{4}} |Y_{\Delta_{l i} }|^2 |Y_{\Delta_{jj}}|^2\,.
\end{equation}
\noindent A signal from MEG~\cite{Kiselev:2009zz} (Table \ref{table:LFV:bounds})
i.e.  $10^{-13} \lesssim{\rm BR}(\mu \to e \gamma) 
\lesssim 10^{-11}$ will put constraints on 
$\Lambda$, depending on the size of the couplings:  assuming 
natural values of $Y_{\Delta}\sim \mathcal{O}(1)$ imply
that 15 TeV $< \Lambda < $ 50 TeV, while for $Y_{\Delta}\sim
\mathcal{O}(10^{-2})$, BR$(\mu \to e \gamma)$ within MEG reach would 
lead to  0.15 TeV $< \Lambda < $ 0.5 TeV.  
\section{cLFV and new physics}
Depending on the mechanism of $\nu$ mass generation, and
especially on the scales of the mediators and size 
of the couplings, one can have very different scenarios of LFV 
at low-energies. 
Whichever NP is called upon to explain the
origin of (Majorana) $\nu$ masses and mixings, the effective
theory will contain at least $d=5$ and $d=6$ operators. 
Whether or not the $d=6$  coefficient is sufficiently 
large to generate observable cLFV strongly depends on two main 
ingredients: the typical scale of
 NP (not necessarily the scale of the seesaw mediator)
and the amount of mixing present in the lepton sector parametrized by an effective mixing $\theta_{i j }$ ($U_{\rm MNS}$ and additionnal mixings).  For $  \mu\to e \gamma$, one approximately has, 
\begin{equation}\label{eq:kuno}
{\rm BR}(\mu \to e\gamma) = 10^{-11}\times (2\ 
{\rm TeV}/\Lambda)^4 
\times (\theta_{\mu e }/0.01)^2\ .
\end{equation}
There are several classes of well-motivated extensions of the SM,
aiming at overcoming both its theoretical and experimental
shortcomings. These models can either offer new explanations for the
smallness of $m_\nu$  (e.g. through a geometrical suppression
mechanism, as is the case of large extraD or R-parity
violation in the case of SUSY models), or onto them one can embed a
seesaw mechanism. In addition, these extensions can provide new
sources of LFV\footnote{It is important to stress that if strictly
  seesaw-related, the new sources of flavour violation will not
  provide additional contributions to hadronic low-energy observables,
  nor to proton decay.}.  
In all cases, once the new fields 
are
integrated out, one finds the $d=5,6$ operators of 
Table~\ref{table:operators}.   
 Notice however that 
in order to firmly establish that a given mechanism is at work,
one needs to observe several cLFV processes that share a 
common source of LFV.

In many of these extensions, the low-energy cLFV observables
can be significantly enhanced when compared to the minimal seesaw~\cite{Raidal:2008jk}.  
As an example, let us consider here the case of a SUSY 
type I seesaw, where cLFV decays are
mediated by sleptons and
gauginos, as illustrated in Figure~\ref{LFVBSM} (c).~
 Flavour violation in the $\nu$ sector is transmitted to the charged
one via radiative effects involving $\nu$ Yukawa couplings $Y^{\nu}$. Even under the
assumption that the SUSY breaking mechanism is flavour conserving,
renormalisation effects (RGE) can induce a sufficiently large  
amount of FV as to account for sizable cLFV 
rates~\cite{Borzumati:1986qx}.
Using the LLog approximation, the dominant
contribution to the  $\ell_i - \ell_j$ transitions  are given by
\begin{equation}\label{eq:BR:MIA:LL}
\hspace*{-8mm}
\frac{
{\rm BR}(\ell_i \to \ell_j \,\gamma)
}{
{\rm BR}(\ell_i \to \ell_j \nu_i\bar \nu_j)
}= 
\frac{\alpha^3\, \tan^2 \beta}{G_F^2\, m_{\rm SUSY}^8}
\left|
\frac{(3\, m_0^2+ A_0^2)}{8\,\pi^2}
\left({Y^{\nu}}^\dagger 
L Y^{\nu}\right)_{ij} 
\right|^2,
\frac{{\rm BR}(\ell_i \to 3 \ell_j)}{{\rm BR}(\ell_i \to \ell_j\, \gamma)}=
 \frac{\alpha}{3\,\pi}
\left(\log\frac{m_{l_i}^2}{m_{l_j}^2}\,-\,\frac{11}{4}\right),
\end{equation} 
where $m_0$ and $A_0$ denote universal soft-SUSY breaking terms,
$m_{\rm SUSY}$ coresponds to an average scale of the SUSY
particles, $\tan \beta$ is the ratio of the Higgs vacuum
expectation values and  having  assumed that 
the dominant contribution to the three body decays originates from $\gamma$-penguin diagrams. 
 
In addition to these low-energy observables, interesting phenomena are
expected to be observed at  colliders. 
In the SUSY seesaw case, one can have sizable widths 
for processes like
$\chi_2^0 \to \chi_1^0 \ell_i^\pm \ell_j^\mp$, flavoured slepton mass
splittings (especially between the first and second generation of
left-handed sleptons) and finally the appearance of new edges in
same-flavour dilepton mass distributions.  
This clearly illustrates that in addition to
directly probing the model of NP by searching for
the new fields at colliders,  
one may also have cLFV at high-energy.  
Furthermore, having a unique source of LFV 
(neutrino mass generation), 
the interplay of  low- and high-energies LFV 
observables can lead to strenghten or disfavour
the underlying model of NP. An illustrative example of the potential  of this interplay can be found for instance in ~\cite{C-A}.

In order to reduce the arbitrariness of the seesaw parameters, 
one can embed the seesaw in GUT frameworks, 
considering larger gauge groups such as SU(5) and SO(10).  
In ordinary non-SUSY SU(5), the addition of a $24_F$ 
fermionic multiplet leads to interesting FV scenarios 
(especially when compared to the inclusion of a $15_H$ Higgs multiplet). 
A mixing of  type I and III seesaws, accompanied 
by the prediction of an SU(2) fermionic triplet below the TeV
scale would lead to a
testable seesaw at the LHC~\cite{Bajc:2006ia}.

cLFV has also been extensively addressed in SUSY GUTs. 
For example,  cLFV observables were analysed
for SO(10)-motivated $Y^\nu$ ans\"atze ~\cite{Calibbi:2006nq} (where at least one
neutrino Yukawa coupling is as large as the top Yukawa
coupling), considering two limits: 
CKM-like and MNS-like mixings. The interplay of the
different observations (or lack thereof) at upcoming facilities would
allow to derive hints on the structure of the unknown neutrino Yukawa
couplings, or even falsify some of the SUSY GUT scenarios.  

Another class of models for generating non-trivial lepton flavour structures
is associated to the displacement of SM fermions 
along extraD (scenarios with either large flat~\cite{ArkaniHamed:1999dc} or 
small warped~\cite{Gherghetta:2000qt} extraDs).  
In the lepton sector, the experimental upper limit on BR$(\mu\to 3 e)$
imposes that the mass of the Kaluza-Klein (KK) excitations be 
$M_{KK} \gtrsim 30$ TeV~\cite{Delgado:1999sv}, thus beyond the reach of the
LHC.  
 
 Finally, cLFV is a feature common to non-SUSY
multi-Higgs doublet extensions of the SM, and can occur both through
charged and neutral currents. In these models, it has been pointed out that even for massless $\nu$ 
lepton flavour can be violated~\cite{B-B}. 
However, in the context of the minimal extension with a type I seesaw,  LFV effects are
extremely small.  
\section{Conclusions}
Up to now, the non-observation of cLFV signals at low-energy has only allowed to derive bounds on models of new physics.  
 Impressive experimental efforts are currently being put in the search for cLFV:  observation of  processes such as $\ell_i \to \ell_j\gamma$, $\ell_i \to \ell_j \ell_k \ell_m$ and  $\mu - e$ conversion would necessarily point towards the existence of heavy states with a typical scale around the TeV, and hence within reach of present colliders (LHC, Tevatron). In addition, one can also expect cLFV in the decay and/or production of these heavy states. 
Different new physics scenarios imply very distinct phenomenology; if cLFV
  is observed, one needs to study the correlations between several
  observables (LFV charged lepton decays and
  LFV at colliders), and to take into account the complementary information 
from new particle searches, in order to be able to pin down the origin of FV.
 Under the assumption of a common source for $\nu$ mass generation and cLFV, the interplay 
 of direct searches for new physics at colliders and indirect signals (such as lepton flavour violating transitions) 
 can hint towards the mechanism of $\nu$-mass generation, probing physics at very high scales (for instance close 
 to the  GUT scale). In particular, from the combination of different observables one might be able  to discriminate between
models of Majorana neutrinos.

\end{document}